\documentstyle[preprint,aps]{revtex}
\tightenlines

\begin{document}
\newcommand{\beq}{\begin{equation}}
\newcommand{\eeq}{\end{equation}}
\newcommand{\beqa}{\begin{eqnarray}}
\newcommand{\eeqa}{\end{eqnarray}}
\newcommand{\sr}{\sqrt}
\newcommand{\fr}{\frac}
\newcommand{\mn}{\mu \nu}
\newcommand{\G}{\Gamma}

\draft
\preprint{ INJE-TP-01-08}
\title{A gauge boson from the Kaluza-Klein approach of the
Randall-Sundrum brane world}
\author{ Y.S. Myung\footnote{E-mail address:
ysmyung@physics.inje.ac.kr} }
\address{
Department of Physics, Graduate School, Inje University,
Kimhae 621-749, Korea}
\maketitle
\begin{abstract}

We clarify a mechanism to obtain a massless gauge boson from the
Kaluza-Klein approach of the Randall-Sundrum(RS) brane world.
This corresponds exactly  to the same mechanism of achieving a
localization of the gauge boson by adding both the bulk and brane
mass terms. Accordingly this work puts another example for
a localization-mechanism of the gauge boson on the brane.

\end{abstract}
\bigskip

\newpage

Recently, there have been lots of interest in the
localization of 4D gravity proposed by Randall and Sundrum (RS)~\cite{RS2,RS1}.
 RS~\cite{RS1} introduced
a single positive tension 3-brane and a negative bulk cosmological
constant in the five-dimensional (5D) spacetime. There have been developed
a large number of brane world models afterwards~\cite{CEHS,OMs}.
The introduction of branes usually gives rise to the ``warping" of the
extra dimensions, resulting in non-factorizable spacetime manifolds.
More importantly, the presence of brane ($\delta(z)$-term) breaks the translational
isometries in the extra dimension and requires  severe boundary
conditions for propagating modes . Therefore, it is a delicate issue to
get a correct information for the propagation of fields in comparison with the
conventional Kaluza-Klein theory.

There are  approaches  to confine standard
particles on the brane by allowing the fields to live in the bulk
spacetime.
In these approaches  it is important to derive the zero mode
effective action because  the zero modes (massless modes) correspond
to the standard model particles localized on the brane.
For example, the zero modes of  bulk scalar and fermion  fields can be localized
on the brane~\cite{GW,Chang}.

On the other hand, the bulk gauge field has a different picture~\cite{BulkFs}. Its zero
mode is not localized on the brane. However, more recently
two interesting models appeared : One is based on the mechanism that the
localization can be achieved by adding both bulk and brane mass
terms\footnote{Authors in\cite{GN} call this the ``inverse-Higgs  mechanism".
 This is so because
two bulk and boundary terms which break the gauge symmetry in the bulk
 are necessary to obtain the
normalizable  wave-function. This broken gauge symmetry in the bulk is restored
on the brane. In the previous works, it is called  a sort of
the ``brane-Higgs effect"\cite{bHiggs,KM}.}~\cite{GN}.
 The other is that the localization of the bulk gauge field
can be realized by taking into account the coupling between the
gauge  and the dilaton field\cite{KT}. Tachibana showed that these two
approaches are closely related to each other\cite{TAC}.

In this paper, we show through the perturbation analysis of the Kaluza-Klein fields
 around the RS background
 that a mechanism that the KK gauge field
possesses the U(1) gauge symmetry on the brane is a kind of the
inverse Higgs mechanism.

Let us start with the second RS model~\cite{RS1,RSmodels}
\beq
I = \int d^4x \int^{\infty}_{-\infty} dz
    \fr{\sr{-\hat{g}}}{16\pi G_5} (\hat{R} -2\Lambda )
    - \int d^4x \sr{-\hat{g}_B} \sigma .
\label{5DI}
\eeq
Here $G_5$ is the 5D Newton's constant, $\Lambda$ the bulk cosmological
constant of 5D spacetime, $\hat{g}_B$ the determinant of the induced
metric describing the brane, and $\sigma$ the tension of the brane.
We consider that the value of $\sigma$ is fine-tuned such that
$\Lambda =-6k^2 (< 0)$ with $k=4\pi G_5 \sigma /3$. Let us introduce
the perturbation around the RS background
\beqa
ds^2 &=& \hat{g}_{MN} dx^Mdx^N = H^{-2}(z) g_{MN} dx^Mdx^N.
\label{metric}
\eeqa
Here $H= k|z|+1(H' =k\theta (z)$, $H''=2k\delta (z))$ is a warp-factor
when one introduces a conformal coordinate
$z$ for the extra dimension instead of $y$.
The standard Kaluza-Klein decomposition of the 5D metric perturbation\footnote{Here we do not
introduce the graviscalar propagation because it
induces an instability problem of the RS background~\cite{ML,Myu,KM}.} is
given by

\beq
(g_{MN}) = \left(\matrix{
\gamma_{\mu\nu}+\kappa^2  a_{\mu}a_{\nu} & -\kappa  a_{\mu}\cr
-\kappa  a_{\nu} & 1\cr}
\right) ,
\qquad
(g^{MN}) = \left(\matrix{
\gamma^{\mu\nu} & \kappa a^{\nu} \cr
\kappa a^{\mu} & (1+\kappa^2 a\cdot a) \cr}
\right)
\label{KKm}
\eeq
with $\gamma_{\mu\nu}=\eta_{\mu\nu}+ \kappa h_{\mu\nu},
\gamma^{\mu\nu}=\eta^{\mu\nu}- \kappa h^{\mu\nu}+\kappa^2 h^{\mu\alpha}h^\nu_\alpha,
a^{\mu} = \gamma^{\mu\alpha} a_{\alpha}= a^{\mu}-\kappa h^{\mu\alpha} a_{\alpha}$
and $a\cdot a =a_{\alpha} a_{\beta} \eta^{\alpha\beta}$.
$\kappa$ is the small parameter for the fluctuation analysis.
Here we keep up to $\kappa^2$-order for our purpose. It is easily
checked that $g_{MP}~g^{NP}=\delta^N_M$ up to this order.

In this work, we are mainly interested in the  bilinear
action of the zero modes (massless modes). In general, it is a non-trivial problem to
determine what the ``zero mode" is if the full spacetime is not
factorizable. As an ansatz for the zero mode, we assume that
$h_{\mu\nu}$ and  $a_{\mu}$ are functions of
$x$-coordinates only : $h_{\mu\nu}(x), a_\mu(x)$.
This  assumption comes from the  observation that
the graviton zero mode $h_{\mn}$ in $\gamma_{\mn}=\eta_{\mn}
+\kappa h_{\mn}$ depends only on ``$x$" even if one starts from
the massive approach of $\hat h_{\mn}(x,z)=H^{3/2} \psi(z) h_{\mn}(x)$ in the RS
model~\cite{RS1}. For the zero mode solution with
$m^2=0$, we have $\psi^0(z)= c_h H^{-3/2}$, thus we find
$\hat h^0_{\mn}(x,z)= c_h  h_{\mn}(x)$ with a constant $c_h$.
For the spin-0 bulk field $\Phi(x,z)=H^{3/2} \chi(z) \phi(x)$, we
have $\chi= c_\Phi H^{-3/2}$ for the zero mode and hence its
localized zero mode is given by $\Phi^0(x,z)= c_\Phi \phi(x)$~\cite{GW}.

Then the five-dimensional action Eq.~(\ref{5DI}) is given by
\beqa
I &=& \fr{1}{16\pi G_5} \int d^4x \sr{-\gamma}
  \Big[(R(\gamma ) -\fr{\kappa^2}{4} f^2)  \int dz H^{-3} +
  (1 +\kappa^2 a\cdot a)
  \int dz H^{-3} \Big( 8\fr{H''}{H} -20\fr{H'^2}{H^2} \Big)
  \nonumber \\
&& +8\kappa \big( \fr{1}{2}
  a^{\mu} \partial_{\mu}h_{\alpha}^{\alpha}
  +\partial_{\mu}a^{\mu} \big) \int dz \fr{H'}{H^4}
  -2\Lambda \int dz H^{-5} \Big]  \nonumber   \\
&& -\int d^4x \sr{-\gamma} \sr{|\delta^{\mu}_{\nu}
  +\kappa^2 a^{\mu}a_{\nu}|} \sigma
\label{FDAC}
\eeqa
with
$f_{\mu\nu}= \partial_{\mu}a_{\nu} -\partial_{\nu}a_{\mu}$.
Considering the relations for the fluctuation analysis around the
Minkowski brane, we have
\beqa
&& R(\gamma)\simeq \delta_1 R(h) +\delta_2 R(h),
\label{EOMg}   \\
&&\sr{|\delta^{\mu}_{\nu} +\kappa^2  a^{\mu}a_{\nu}|}
\simeq 1+\fr{1}{2} \kappa^2  a\cdot a,
\label{EOMp} \\
&& \sqrt{-\gamma} \simeq 1 + \fr{1}{2} h_\alpha^\alpha -\fr{1}{4}(
h_{\alpha}^{\beta}h^{\alpha}_{\beta}-
  \fr{1}{2}h_{\alpha}^{\alpha}h_{\beta}^{\beta}),
\label{EOMr}
\eeqa
where $\delta_1 R(h) $ and $\delta_2 R(h)$ are the linear and
bilinear Ricci scalar terms, respectively.
Then the bilinear action to  Eq.~(\ref{FDAC})  which
governs the perturbative dynamics is given by

\beqa
I_{bilinear} &=& \fr{\kappa^2}{16\pi G_4} \int d^4x \int dz \Big\{-
\fr{1}{4H^3}(
  \partial^{\mu}h^{\alpha\beta}\partial_{\mu}h_{\alpha\beta}
  -\partial^{\mu}h\partial_{\mu}h +2\partial^{\mu}h_{\mu\nu}
  \partial^{\nu}h -2\partial^{\mu}h_{\mu\alpha}\partial^{\nu}
  {h_{\nu}}^{\alpha} )  \nonumber  \\
&& \qquad\qquad -\fr{1}{4H^3}(\partial_{\mu}a_{\nu} -\partial_{\nu}a_{\mu})
  (\partial^{\mu}a^{\nu} -\partial^{\nu}a^{\mu}) \nonumber \\
&&
  -\fr{1}{H^5}(-2k^2 + k\delta(z))(h_{\alpha}^{\beta}h^{\alpha}_{\beta}-
  \fr{1}{2}h_{\alpha}^{\alpha}h_{\beta}^{\beta})+
  \fr{10}{H^5}(-2k^2+ k\delta(z))a^{\mu}a_{\mu}\Big\}
\label{KKILP}
\eeqa
up to the partial integration over $d^4x$. Interestingly,
it turns out that the terms in the last line of Eq.(\ref{KKILP})
look like the mass terms. $-2k^2$-terms come from the bulk
AdS information of $\Lambda=-6k^2$, whereas $k\delta(z)$-terms arise
from the presence of the brane at $z=0$. In the case of bulk gauge
boson, one inserts both the bulk and brane mass-terms by hand.
Here the combined effect of the Kaluza-Klein approach with the brane
world scenario makes the same thing as in the addition of the bulk
and boundary mass-terms in the bulk gauge field action to obtain
the normalizable wave function~\cite{GN}.

In order to see what physical states there are, let us analyze
the field equations as below. First we wish to do it without any integration
over $z$.  From the action Eq.~(\ref{KKILP}) we have
the equations of motion
\beqa
\mbox{} && \fr{1}{H^3} \Big[\Box h_{\mu\nu} +\partial_{\mu}\partial_{\nu}
  h -( \partial_{\mu}\partial^{\alpha}h_{\alpha\nu}
  +\partial_{\nu}\partial^{\alpha}h_{\alpha\mu} )
  -\eta_{\mu\nu} ( \Box h -\partial^{\alpha}\partial^{\beta}
  h_{\alpha\beta} ) \Big]  \nonumber \\
&& = \fr{4k}{H^5}(2k-\delta(z))(\fr{h}{2}\eta_{\mu\nu}-h_{\mu\nu}),
\label{vspin2}  \\
&& \Box a_{\mu} -\partial_{\mu} (\partial_{\nu}a^{\nu})
=\fr{40k}{H^5}(2k-\delta(z))a_\mu.
\label{vspin1}
\eeqa
Here we find that the right hand terms of Eqs.(\ref{vspin2}) and
(\ref{vspin1}) look like the  mass-terms  for graviton and graviscalar.
Especially  the presence of the brane at $z=0$ gives rise to the singular
behaviors and thus it requires the  boundary conditions
on $h_{\mu\nu}$ and $a_\mu$ along the extra direction.
Hence as they stand, these are not genuine massless spin 2 and spin 1 particles
on the brane.
Our goal is to find massless particles.

In order to obtain  a truly propagating graviton and a KK gauge
boson, we have to integrate these equation over $z$ using
 $\int^{\infty}_{-\infty}
dz \delta(z)=1$, $\int^{\infty}_{-\infty}
dz H^{-3} = 1/k$, and $\int^{\infty}_{-\infty} dz H^{-5} = 1/2k$.
This corresponds to the fine-tuning  to obtain massless modes in the linearized equation.
Then the right hand terms of Eqs.(\ref{vspin2}) and (\ref{vspin1})
vanish identically and thus the massless
modes appear as~\cite{KM}
\beqa
\mbox{} && \Box h_{\mu\nu} +\partial_{\mu}\partial_{\nu}
  h -( \partial_{\mu}\partial^{\alpha}h_{\alpha\nu}
  +\partial_{\nu}\partial^{\alpha}h_{\alpha\mu} )
  -\eta_{\mu\nu} ( \Box h -\partial^{\alpha}\partial^{\beta}
  h_{\alpha\beta} )=0,
\label{spin2}  \\
&& \Box a_{\mu} -\partial_{\mu} (\partial_{\nu}a^{\nu})
=0.
\label{spin1}
\eeqa
 In other words, in the bilinear action Eq.(\ref{KKILP}),
the condition that the massless modes
are localized on the brane requires  the finiteness of its
integral after the integration over $z$. In this case the
mass-like terms of $a^\mu a_\mu$ and $ h_{\alpha}^{\beta}h^{\alpha}_{\beta}-
  \fr{1}{2}h_{\alpha}^{\alpha}h_{\beta}^{\beta}$ identically
  disappear too. Then we can keep the U(1) gauge symmetry in the
  bilinear action. This is
   the brane-Higgs effect in our KK gauge field approach. Further this can be
   interpreted as  the inverse-Higgs mechanism
   in the bulk gauge field approach.
Now we take the trace of Eq.~(\ref{spin2}) to get a constraint
\beq
\Box h -\partial^{\alpha}\partial^{\beta} h_{\alpha\beta}
  = 0  .
\label{Trace}
\eeq
Hence Eq.~(\ref{spin2}) becomes
\beq
\Box h_{\mu\nu} +\partial_{\mu}\partial_{\nu}
  h -\Big( \partial_{\mu}\partial^{\alpha}h_{\alpha\nu}
  +\partial_{\nu}\partial^{\alpha}h_{\alpha\mu} \Big)
  = 0 .
\label{spin2b}
\eeq

So far we have not chosen any gauge for $h_{\mu\nu}$. Now let us
choose the transverse (or harmonic) gauge in the five-dimensional
spacetime. Considering

\beq
g_{MN} \simeq \eta_{MN} +\kappa \epsilon_{MN}, \qquad\qquad
\Big( \epsilon_{MN} \Big) = \left(\matrix{
h_{\mu\nu} & -a_{\mu} \cr
-a_{\nu} & 0 \cr}
\right)  ,
\eeq
the five-dimensional harmonic gauge $\partial^M \epsilon_{MN}
=\fr{1}{2} \partial_N \epsilon$ is equivalent to
\beq
\partial^{\mu}h_{\mu\nu} =\fr{1}{2}\partial_{\nu} h,
\qquad\qquad  \partial_{\mu}a^{\mu} =0.
\eeq
This means that the 5D harmonic gauge is split into the harmonic gauge for
the 4D gravitational field and the Lorentz gauge for the 4D KK gauge
field. Using these gauge conditions above, Eq.~(\ref{spin2b}) and
Eq.~(\ref{spin1}) become
\beq
\Box h_{\mu\nu} = 0 , \qquad\qquad\qquad
\Box a_{\mu} = 0 ,
\label{spin2c}
\eeq
respectively. Therefore, it proves that $h_{\mu\nu}$ and
$a_{\mu}$ indeed represent the massless spin-2 particle (graviton) and the
KK massless spin-1 particle (gauge boson) on the brane, respectively.

In studying the  U(1) Maxwell term  arisen from the 5D RS brane model, we use
 both the conventional Kaluza-Klein approach and the fluctuation analysis.
  Apparently, we find a mass-like term $ k^2 a\cdot a$ from the bulk AdS space
   as well as a (mass-like) singular term
 $ k \delta(z) a\cdot a$ from the presence of the brane at $z=0$.
 In order to obtain the genuine massless vector, we integrate it
 over $z$. Then these terms disappear in the linearized equation
 (or equivalently, the vanishing of the mass-like terms in the
 bilinear action). Previously we interpreted it as  a sort of
the brane-Higgs effect~\cite{bHiggs,KM}: The isometry of extra dimension was broken
spontaneously by the presence of the brane. Hence we expect that
the gauge field becomes massive. However, we have found after the integration over $z$
that the
massive propagation of the KK gauge field does not reveal at the linear
level (or in the bilinear action).  We find the massless vector propagation.
Here the procedure of the integration over $z$ is a crucial step for  obtaining
the KK massless gauge boson. This actually corresponds to the
procedure of obtaining  localized zero modes of graviton and scalar in the
bulk approach.

Finally we have a few comments in order. The first important one is that our
``brane-Higgs effect" for the KK gauge field corresponds to the
``inverse-Higgs mechanism" in the bulk gauge field approach of
introducing two bulk and boundary mass terms~\cite{GN}. Our approach is more
natural than the the inverse-Higgs mechanism because our KK
setting with the brane gives us the massless gauge boson, whereas
in the latter case one has to introduce two mass terms by hand to obtain the massless
gauge boson on the brane. But the results are the same.
The second is that the procedure of obtaining the gauge boson
bears
a close parallel to that of the 4D massless graviton.
Because the zero mode sector for the 4D graviton in the second RS model was well
established~\cite{RS1}, our result for the zero mode for 4D KK gauge field is
very credible.
The last one concerns about the non-existence argument of the
gauge boson in the Randall-Sundrum model. This is based on the the
Z$_2$ orbifold symmetry such that the vector gauge field $A_\mu$ should
satisfy $A_\mu(x,-z)=-A_\mu(x,z)$~\cite{RS2}. So we have $A_\mu(x,0)=0$.
Hence if one requires the Z$_2$ orbifold symmetry in the brane
world model, there will be no vector zero mode fluctuations.
This may be true before the integration of the bilinear action over
$z$. However, if we accept the fact that  the last condition for
obtaining a gauge boson is to integrate the linearized equation
(or, the bilinear action) over $z$,
 we find the massless
gauge boson on the brane as an outcome of the brane-Higgs effect.

\section*{Acknowledgments}

The authors thank H.W. Lee and Gungwon Kang for helpful discussions.
 This work was supported by the Brain Korea 21 Programme,
Ministry of Education, Project No. D-1123.

\end{document}